\documentstyle[12pt]{article}

\newcommand{\spone}{0.9}

\newcommand{\spthree}{2.4}

\newcommand{\singlespace}{\edef\baselinestretch{\spone}\Large\normalsize}

\newcommand{\threespace}{\edef\baselinestretch{\spthree}\Large\normalsize}
\threespace
\begin{document}
\singlespace

\begin{center}
{\bf Stability of Solution of the Nonlinear Schr\"odinger Equation
for the Bose-Einstein Condensation} \\
\vspace{8pt}

{\bf Yeong E. Kim and Alexander L. Zubarev} \\
\vspace{5pt}

Department of Physics, Purdue University \\
West Lafayette, IN  47907-1397 \\
\vspace{12pt}

\end{center}
\vspace{12pt}

We investigate the stability of the Bose-Einstein condensate (BEC) for the
case of atoms with negative scattering lengths at zero temperature using
the Ginzburg-Pitaevskii-Gross (GPG) stationary theory.  We have found
a new exact equation for determining the upper bound
of the critical numbers $N_{cr}$ of atoms for a metastable state to exist.
Our calculated value of $N_{cr}$ for Bose-Einstein condensation of lithium
atoms based on our new equation is in agreement with those observed in a recent experiment.
\vspace{24pt}

\noindent
PACS number(s):  03.75.Fi, 03.65.Db, 05.30.Jp, 67.90.+z
\vspace{24pt}

\noindent
{\bf 1.  Introduction}
\vspace{5pt}

The concept of the Bose-Einstein condensation (BEC) [1] has been known
for 73 years, and has been used to describe all physical scales, including
liquid $^4He$, exitons in semiconductors, pions and kaons in dense nuclear
matter (neutron stars, supernovae), and elementary particles [2].  It is only
a few years ago that the BEC phenomenon was observed directly in dilute
vapors of alkali atoms, such as rubidium [3], lithium [4,5], and sodium [6],
confined in magnetic trap and cooled down to nanokelvin temperatures.  At
such low temperatures, the thermal de Brogle wavelength increase to submicrometer
dimensions.  When the scattering length is negative, then nonlinear
interaction between atoms is attractive and it has been claimed that the
BEC in free space is impossible [7], because the attraction makes the system
tend to an ever denser phase.  For $^7Li$, the s-wave scattering length
${\it a}$ = (-14.5 $\pm$ 0.4) $\AA$ [8].  For bosons trapped in an external
potential, there may exist a metastable BEC state under certain conditions
for a number of atoms below the critical value $N_{cr}$ [9-11]. Stability of the
BEC in a trap as a function of nonlinearity has been studied numerically for
T = 0 [12 - 14].  It was demonstrated [15, 16] that global solutions of
the Ginzburg-Pitaevskii-Gross equation (GPG) [17] for trapped atoms with
negative energies diverge in a finite time.  When the system collapses, a
more accurate theory is required in order to include short range effects.
\vspace{5pt}

In this paper, we report our theoretical studies of the stability of the BEC for the case with negative scattering
length at zero temperature using the GPG stationary theory [17].  We
have found, for the first time, an exact equation for determination of the
upper bound of the $N_{cr}$ for a metastable state to exist.
\vspace{8pt}

\noindent
{\bf 2.  A necessary condition for a local minimum}
\vspace{5pt}

In the mean-field approximation, the ground state energy of the system
is given by the GPG energy functional [17]
$$ J(\psi) = \bar{T} + \bar{V} + \frac{g_oN}{2} \bar{V}_H ,
\eqno{(1)}
$$

\noindent
with
$$ J(\psi) = E/N ,
\eqno{(2)}
$$

$$\bar{T} = \int~d^3r~\frac{\hbar^2}{2m}|\nabla \psi|^2 ,
\eqno{(3)}
$$

$$\bar{V} = \int~d^3r~\psi^\ast V \psi ,
\eqno{(4)}
$$

$$\bar{V}_H = \int~|\psi|^4d^3r ,
\eqno{(5)}
$$

\noindent
and

$$ g_o = \frac{4 \pi \hbar^2 {\it a}}{m} ,
\eqno{(6)}
$$

\noindent
where {\it a} is the s-wave scattering length, N is the number of particles in the BEC, V is an external trapping potential, and $\psi$ is the condensate ground
state wave function normalized as 
$$ \int|\psi|^2d^3r = 1 .
\eqno{(7)}
$$

\noindent
For a harmonic oscillator trap, we have
$$ V(\vec{r}) = V_1(x)+V_2(y)+V_3(z),
\eqno{(8)}
$$

\noindent
where

$$ V_i(r_i) = \frac{m}{2} \omega_i^{(o)2}r_i^2.
$$

\noindent
Using
$$V_i(\lambda_i r_i) = \lambda_i^2 V_i(r_i) ,
\eqno{(9)}
$$

\noindent
we can obtain an extremum condition (see Appendix)
$$ 2 \bar{T}_i - 2 \bar{V}_i + \frac{g_oN}{2} \bar{V}_H = 0 ,
\eqno{(10)}
$$

$$ \bar{T}_i = \int d^3r \frac{\hbar^2}{2m}|\nabla_i \psi|^2 ,
\eqno{(11)}
$$

$$ \bar{V}_i = \frac{m}{2} \int d^3r \psi^\ast(\vec{r})(\omega_i^{(o)2}r_i^2)
\psi(\vec{r}) ,
\eqno{(12)}
$$

\noindent
and a necessary local minimum condition
$$ \hat{A} > 0 ,
\eqno{(13)}
$$

\noindent
where $\hat{A}$ is matrix with matrix elements
$$ A_{ij} = (2 \bar{T}_i + 6 \bar{V}_i)\delta_{ij} + (1 - \delta_{ij})
\frac{g_oN}{2} \bar{V}_H  .
\eqno{(14)}
$$

\noindent
(See Appendix for details.)
\vspace{8pt}

Eqs. (10-12) represent  well-known virial theorem equations [12], but the
conditions (13) was not discussed in literature, to the best of our
knowledge. We note, that if the conditions (13) are not fulfilled, local
minimum disappears and solution will be not stable.  In the case
${\it a} <$ 0, we can rewrite (14) as
$$ A_{ij} = \frac{|g_o|N\bar{V}_H}{2}[\tilde{g}_i \delta_{ij} +
2\delta_{ij} - 1],
\eqno{(15)}
$$

\noindent
where
$$ \tilde{g}_i = \frac{16 \bar{V}_i}{|g_o|N\bar{V}_H}
\eqno{(16)}
$$

\noindent
Matrix ($2 \delta_{ij} - 1$) is not positive (its eigenvaliues are
-1, 2, 2) hence in case $V_1 = V_2 = V_3 = 0$ (free space) matrix
A, is not positive and there is no stable solution.
\vspace{8pt}

\noindent
{\bf 3. Upper bounds for $N_{cr}$.}
\vspace{5pt}

We consider first an isotropic trap case with
$$ \tilde{g}_x = \tilde{g}_y = \tilde{g}_z = \frac{16}{3} \frac{\bar{V}}{|g_o|N\bar{V}_H} ,
\eqno{(17)}
$$

\noindent
where

$$ \bar{V} = \bar{V}_x + \bar{V}_y + \bar{V}_z = 3 \bar{V}_x .
\eqno{(18)}
$$

\noindent
For this case the local minimum conditions given by Eq. (13) can be written as
$$ \frac{16 \bar{V}}{3|g_o|N\bar{V}_H} > 1. 
\eqno{(19)}
$$

\noindent
In terms of $\bar{N}$, which is solution of the following equation
$$ \bar{N} = \frac{16}{3} \frac{\bar{V}}{|g_o| \bar{V}_n} ,
\eqno{(20)}
$$

\noindent
we can state that if $N \geq \bar{N}$, the BEC is not stable. Therefore
$\bar{N}$ is an upper bound for $N_{cr}$ ,
$$ N_{cr} \leq \bar{N} .
\eqno{(21)}
$$

\noindent
We note, that Eq.(21) is also valid for anisotropic cases, but the
bound given by Eq. (20) in this case can be improved.  To show this let us consider
an anisotropic harmonics trap
$$ V(\vec{r}) = V_\bot(\vec{r}) + V_z(\vec{r}) ,
\eqno{(22)}
$$

\noindent
here
$$ V_\bot(\vec{r}) = \frac{m}{2} \omega_\bot^{(o)2}(x^2 + y^2) ,
\eqno{(23)}
$$

$$ V_z(\vec{r}) = \frac{m}{2} \omega_z^{(o)2} z^2 .
\eqno{(24)}
$$

Using Eqs. (22-24) we can rewrite the local minimum conditions (13) as
$$ \frac{4 \bar{V}_\bot}{|g_o|N \bar{V}_H} + (\frac{8}{|g_o|N \bar{V}_H})^2~
\bar{V}_\bot~\bar{V}_z - 1 > 0 ,
\eqno{(25)}
$$

\noindent
where
$$ \bar{V}_\bot = \frac{m}{2} \omega_\bot^{(o)2}~\int~\psi^\ast(\vec{r})
  (x^2 + y^2)\psi(\vec{r})d^3\vec{r} ,
\eqno{(26)}
$$

$$\bar{V}_z = \frac{m}{2} \omega_z^{(o)2}~\int~\psi^\ast(\vec{r})z^2\psi(\vec{r})d^3\vec{r} .
\eqno{(27)}
$$

\noindent
Setting the left-hand side of Eq. (25) to zero we obtain the following
equation
$$ \bar{N} = \frac{2 \bar{V}_\bot}{|g_o| \bar{V}_H} [1 + (1 + 
\frac{16 \bar{V}_z}{\bar{V}_\bot})^{1/2}]
\eqno{(28)}
$$

\noindent 
for upper bound $\bar{N}$
$$ N_{cr} \leq \bar{N} .
\eqno{(29)}
$$

\noindent
It should be pointed out that our formulas (28) and (29) are valid for
a general anisotropic trap with
$$ \bar{V}_\bot =\frac{m}{2} \int \psi^\ast (\vec{r})(\omega_x^{(o)2} x^2 +
\omega_y^{(o)2} y^2)\psi(\vec{r})d\vec{r} .
\eqno{(30)} 
$$

To the best of our knowledge, these formulas have never been discussed
previously in the literature.
\vspace{8pt}

\noindent
{\bf 4.  Application to lithium atoms:  numerical example}
\vspace{5pt}

We assume the trial wave function for $\psi$ in Eqs. (10-28) to be [14, 18]
$$ \psi_t(\vec{r}) = \omega_\bot^{1/2} \omega_z^{1/2} (\frac{m}{\pi \hbar})^{3/4}
e^{-m(\omega_\bot^2(x^2 + y^2) + \omega_z^2z^2)/2\hbar}.
\eqno{(31)}
$$

\noindent
Using Eq. (31), we rewrite Eqs. (10) and (28) as
$$ 1 - 2\delta_\bot^2 = \delta_\bot^2 (1 + 8 \frac{\delta_z}{\delta_\bot} \lambda^2)^{1/2} ,
\eqno{(32)}
$$

$$1 - \lambda^2 \delta_z^2 = \delta_z\delta_\bot[1 + (1 + 8 \frac{\delta_z}{\delta_\bot} \lambda^2)^{1/2}],
\eqno{(33)}
$$

\noindent
and
$$n = \delta_z\delta_\bot[1 + (1 + 8 \frac{\delta_z}{\delta_\bot} \lambda^2)^{1/2}] ,
\eqno{(34)}
$$

\noindent
where $\delta_z = \omega_\bot^{(o)}/\omega_z, \delta_\bot = \omega_\bot^{(o)}/\omega_\bot, \lambda = \omega_z^{(o)}/\omega_\bot^{(o)}$, and
$$n = 2(\omega_\bot^{(o)})^{1/2}~(\frac{m}{2 \pi \hbar})^{1/2}~\bar{N}|{\it a}|.
\eqno{(35)}
$$

We find that numerical solution of Eqs. (32-34) for $0 \leq \lambda \leq 1$ can be
interpolate as 
$$ n = e^{-(\alpha + \beta \lambda^2)}
\eqno{(36)}
$$

\noindent
with $\alpha$ = 0.490419, $\beta$ = 0.149175.  Using Eqs. (36) and (35) we
have
$$ \bar{N} = (\frac{2 \pi \hbar}{\omega_\bot^{(o)}m})^{1/2}~~
\frac{e^{-(\alpha + \beta \lambda^2)}}{2|{\it a}|}
\eqno{(37)}
$$

\noindent
Taking the experimental parameters [5], $\omega_\bot^{(o)}/2 \pi$ = 152 Hz, and
$\omega_z^{(o)}/2 \pi$ = 132 Hz, we obtain $\lambda$ = 0.86842 and 
 $ N_{cr} \leq \bar{N}$= 1456.
This value of $ N_{cr}$ is consistent with theoretical predictions [12-14] and is in
agreement with those observed in a recent experiment [5].
\vspace{8pt}

\noindent
{\bf 5. Summary and Conclusions}
\vspace{5pt}

We have investigated the stability of the BEC for the case of
atoms with  negative scattering
length at zero temperature in a magnetically trapped atomic gas.
Using a rigorous derivation, we have found a new exact equation for determining
the upper bound of the critical number of atoms $N_{cr}$ for a metastable
state to exist.  Our calculated value of $N_{cr}$ for BEC  of lithium is
about 1456 atoms, which is consistent with recent experimental measurements
[5] and theoretical predictions [12-14].
\pagebreak

\begin{center}
{\bf Appendix}
\end{center}
\vspace{5pt}

In this appendix, we give proving Eqs. (10-14).  Let us introduce a
trial function
$$ \psi_t(\vec{r}) = [(1 + \epsilon_1)(1 + \epsilon_2)(1 + \epsilon_3)]^{1/2}~
\psi((1 + \epsilon_1)x, (1 + \epsilon_2)y, (1 + \epsilon_3)z),
\eqno{(A.1)}
$$

\noindent
where $\psi(x, y, z)$ is exact ground state solution of the GPG equation.
The evaluation of $J(\psi_t)$, Eq. (1), can be carried out as
$$\everymath={\displaystyle}
\begin{array}{rcl}
J(\psi_t) &\equiv& J(\epsilon_1, \epsilon_2, \epsilon_3) = (1 + \epsilon_1)^2
\bar{T}_x + (1 + \epsilon_2)^2 \bar{T}_y + (1 + \epsilon_3)^2\bar{T}_z + \\
+ \frac{1}{(1 + \epsilon_1)^2} \bar{V}_x &+& \frac{1}{(1 + \epsilon_2)^2} \bar{V}_y
+ \frac{1}{(1 + \epsilon_3)^2} \bar{V}_z +(1 + \epsilon_1)(1 + \epsilon_2)(1 +
\epsilon_3) \frac{g_oN}{2} \bar{V}_H ,
\end{array}
\eqno{(A.2)}
$$

\noindent
where
$$\everymath={\displaystyle}
\begin{array}{rcl}
\bar{T}_i &=& \int~d^3r~\frac{\hbar^2}{2m} |\nabla_i\psi|^2,
\end{array}
$$

\noindent
and
$$\everymath={\displaystyle}
\begin{array}{rcl} 
\bar{V}_i &=& \frac{m}{2} \int d^3r \psi^\ast(\vec{r})(\omega_i^{(o)2}r_i^2)\psi(\vec{r})  .
\end{array}
$$

\noindent
Since $\psi(\vec{r})$ is exact ground state solutions, $J(\epsilon_1, \epsilon_2,
\epsilon_3)$ has an extremum at $\epsilon_1= \epsilon_2= \epsilon_3 = 0$, which
requires that
$$ \frac{\partial J(\epsilon_1, \epsilon_2, \epsilon_3)}{\partial \epsilon_i}\Biggr\vert_{\epsilon_1 = \epsilon_2 = \epsilon_3 = 0} = 0  .
\eqno{(A.3)}
$$

\noindent
Substitution (A.2) into (A.3) yields Eq. (10),
$$ 2 \bar{T}_i - 2 \bar{V}_i + \frac{g_oN}{2}~\bar{V}_H = 0.
\eqno{(A.4)}
$$

\noindent
Eq. (A.4) represents well-known virial theorem equations [12].
\vspace{8pt}

Now, let us introduce matrix $\hat{A}$ with matrix elements
$$ A_{ij} = \frac{\partial^2 J(\epsilon_1, \epsilon_2, \epsilon_3)}{\partial \epsilon_i \partial \epsilon_j}\Biggr\vert_{\epsilon_1, \epsilon_2, \epsilon_3 = 0} =
(2 \bar{T}_i + 6 \bar{V}_i)\delta_{ij} + (1 - \delta_{ij}) \frac{g_oN}{2}
\bar{V}_H .
\eqno{(A.5)}
$$

\noindent
In terms of $\hat{A}$ we can write a local minimum conditions as
$$ \hat{A} > 0
\eqno{(A.6)}
$$
\pagebreak

\begin{center}
{\bf References}
\end{center}
\vspace{8pt}

\noindent
[1]  S. N. Bose, Z. Phys. {\bf 26}, 178 (1924); A. Einstein, Sitz. Press
Akad. Wiss. {\bf 1924}, 3 (l924). 
\vspace{5pt}

\noindent 
[2]  A. Griffin, D. Snoke, and S. Stringari, {\it Bose-Einstein Condensation}
(Cambridge, New York, 1995).
\vspace{5pt}

\noindent
[3]  M. H. Anderson, J. R. Ensher, M. R. Matthews, C. E. Wieman, and
E. A. Cornell, Science {\bf 269}, 198 (1995).
\vspace{5pt}

\noindent
[4]  C. C. Bradley, C. A. Sackett, J. J. Tollett, and R. G. Hulet, Phys.
Rev. Lett.. {\bf 75}, 1687 (1995).
\vspace{5pt}

\noindent
[5]  C. C. Bradley, C. A. Sackett, and R. G. Hulet, Phys. Rev. Lett.
{\bf 78}, 985 (1997).
\vspace{5pt}

\noindent
[6] K. B. Davis, M.-O. Mewes, M. R. Andrew, N. J. Van Druten, D. S.
Durfes, D. M. Kurn, and W. Ketterle, Phys. Rev. Lett. {\bf 75}, 3969 (1995).
\vspace{5pt}

\noindent
[7]  T. D. Lee, K. Huang, and C. N. Yang, Phys. Rev. {\bf 106}, 1135 (1957).
\vspace{5pt}

\noindent
[8]  E. R. I. Abraham, W. I. McAlexander, C. A. Sackett, and R. G. Hulet,
Phys. Rev. Lett. {\bf 74}, 1315 (1995).
\vspace{5pt}

\noindent
[9]  Y. Kagan, G. V. Shlyapnikov, and J. T. M. Walraven, Phys. Rev. Lett.
{\bf 76}, 2670 (1996).
\vspace{5pt}

\noindent
[10] H. T. C. Stoof, J. Stat. Phys. {\bf 87}, 1353 (1997).   
\vspace{5pt}

\noindent
[11] E. V. Shuryak, Phys. Rev. A {\bf 54}, 3151 (1996).
\vspace{5pt}

\noindent
[12] F. Dalfovo and Stringari, Phys. Rev. {\bf A53}, 2477 (1996).  
\vspace{5pt}

\noindent
[13] R. J. Dodd, M. Edwards, C. J. Williams, C. W. Clark, M. J. Holland, 
P. A. Ruprecht, and K. Burnett, Phys. Rev. {\bf A54}, 661 (1996).
\vspace{5pt}

\noindent
[14] Hualin Shi and Wei-Mou Zheng, Phys. Rev. {\bf A55}, 2930 (1997). 
\vspace{5pt}

\noindent
[15] L. P. Pitaevskii, Report cond.-mat./9605119. 
\vspace{5pt}

\noindent
[16] Kimitaka Watanabe, Tetsuga Mukai, and Takaaki Mukai, Phys. Rev.
{\bf A55}, 3639 (1997).
\vspace{5pt}

\noindent
[17] L. Ginzburg, and L. P. Pitaevskii, Zh. Eksp. Teor. Fiz. {\bf 34}, 1240
(1958) [Sov. Phys. JETP 7, 858 (1958)]; E. P. Gross, J. Math. Phys. {\bf 4},
195 (1963).
\vspace{2pt}

\noindent
[18] G. Baym and C. J. Pethick, Phys. Rev. Lett. {\bf 76}, 6 (1996).

\end{document}